\documentclass[]{article}
\usepackage{emulateapj,onecolfloat,psfig}

\newcommand{\be}{\begin{equation}}
\newcommand{\ee}{\end{equation}}

\def\spose#1{\hbox to 0pt{#1\hss}} 
\def\ltsim{\mathrel{\spose{\lower.5ex\hbox{$\mathchar"218$}}
     \raise.4ex\hbox{$\mathchar"13C$}}}
\def\gtsim{\mathrel{\spose{\lower.5ex \hbox{$\mathchar"218$}}
     \raise.4ex\hbox{$\mathchar"13E$}}}

\def\eg{{\it e.g.}}
\def\etal{{\it et al.}}

\def\ie{{\it i.e.}}
\slugcomment{\it To appear in ApJ}

\begin{document}

\twocolumn
[
\title{Bars and Dark Matter Halo Cores}
\author{J. A. Sellwood}
\affil{Rutgers University, Department of Physics \& Astronomy, \\
       136 Frelinghuysen Road, Piscataway, NJ 08854-8019 \\
       {\it sellwood@physics.rutgers.edu}}

\begin{abstract}
Self-consistent bars that form in galaxies embedded within cuspy halos are 
unable to flatten the cusp.  Short bars form in models with quasi-flat rotation 
curves.  They lose angular momentum to the halo through dynamical friction, but 
the continuous concentration of mass within the disk as the bar grows actually 
compresses the halo further, overwhelming any density reduction due to the 
modest angular momentum transfer to the halo.  Thus the Weinberg-Katz proposed 
solution to the non-existence of the predicted cuspy halos from CDM simulations 
would seem to be unworkable.  I also find that the concerns over the performance 
of $N$-body codes raised by these authors do not apply to the methods used here.
\end{abstract}

\keywords{galaxies: formation --- galaxies: kinematics dynamics} ]

\section{Introduction}
It now seems that the cuspy density profile of halos which form in simulations 
of the dark matter (DM) component is a robust and reproducible feature of the 
collisionless dynamics of halo formation (\eg\ Power \etal\ 2002).  The precise 
slope of the inner profile, if it can indeed be characterized by a power law, is 
still disputed, but all agree that the DM density rises steeply towards the 
center.

Tests of this prediction require estimates of the DM density profiles in 
galaxies.  In principle, the mass profile of DM can be deduced from rotation 
curve measurements, once the contribution due to the observed baryonic material 
has been subtracted.  Attention has generally focused on DM dominated galaxies, 
such as low surface brightness (LSB) or dwarf galaxies, because of their 
believed smaller relative baryonic content today.

There is a substantial and increasing body of high-resolution optical data to 
indicate that small and LSB galaxies have slowly rising rotation curves (\eg\ 
Rubin \etal\ 1985; Courteau 1997; Palunas \& Williams 2000; Blais-Ouellette 
\etal\ 2001; de Blok \etal\ 2001; Matthews \& Gallagher 2002) indicating that 
their DM halos today have rather uniform density cores irrespective of the 
stellar contribution.

Estimates of the DM density profile in larger galaxies are more difficult 
because of the relatively greater baryonic contribution to the rotation curve. 
Englmaier \& Gerhard (1999), Weiner \etal\ (2001), and Kranz, Slyz \& Rix (2002) 
have made {\it dynamical\/} estimates of the disk mass, finding generally that 
the disk dominates the inner galaxy and the DM halo has a very low central 
density.  Binney \& Evans (2001) also argue, though slightly less directly, for 
a low central density in the Milky Way.

A much less direct, but more general, argument was advanced by Debattista \& 
Sellwood (1998, 2000).  The fact that bars appear not to have been slowed by 
friction places an upper bound to the central DM density in barred galaxies.  
The dimensionless parameter ${\cal R} = D_{\rm L}/a_{\rm B}$, with $D_{\rm L}$ 
is the radius of corotation and $a_{\rm B}$ the semi-major axis of the bar, is 
measured to lie in the range $1 \ltsim {\cal R} \ltsim 1.4$ in several cases 
(Aguerri \etal\ 2002), while other, less direct, evidence also suggests that 
bars end only slightly inside their corotation circles.  Debattista \& Sellwood 
found that such low values of $\cal R$ can be sustained only if the halo has a 
low central density.

Low central densities of DM in galaxies today need not be a problem for the 
currently favored $\Lambda$CDM model if the cusps could be subsequently erased 
during galaxy formation or evolution.  At least three ideas to lower the central 
DM density have been proposed:

\begin{itemize}
\item Binney, Gerhard \& Silk (2001), and others have proposed that the halo 
profile is altered by adiabatic compression as the gas cools followed by 
impulsive outflow of a large fraction of the baryon mass.  One possible 
mechanism to produce such an outflow might be a burst of star formation.  The 
idea was examined by Navarro, Eke \& Frenk (1996) and by Gnedin \& Zhao (2002), 
who found that only a mild reduction in the central DM density could be achieved 
in this way.  Gnedin \& Zhao tested the extreme case that 100\% of the baryonic 
component was somehow blasted out instantaneously, yet found that even in this 
deliberately unrealistic case, the central density decreased by little more than 
a factor 2, unless the initial baryons were unrealistically concentrated to the 
halo center. 

\item Milosavljevi\'c \etal\ (2002) point out that a binary supermassive black 
hole (BH) pair created from the merger of two galaxy fragments will eject DM 
(and stars) from the center of the merger remnant.  They also argue that the DM 
mass removed for a given final BH mass is greater if the final BH is built up in 
a series of mergers each having correspondingly lower mass BHs.  While this 
mechanism must operate wherever binary BHs have been formed, the radial extent 
over which the mass is reduced is rather limited (typically a few hundred pc), 
whereas the radial extent of DM cores in brighter galaxies (\eg\ Weiner \etal\ 
2001) is too large to be accounted for by this mechanism alone.  Furthermore, 
cores are observed in DM-dominated galaxies with insignificant bulges which are 
likely to have very low-mass BHs (Gebhardt \etal\ 2001; Ferrarese \& Merritt 
2001), if they contain BHs at all.

\item El-Zant \etal\ (2001) propose that the cusp in the halo density can be 
erased by dynamical friction with orbiting mass clumps.  Weinberg \& Katz (2001) 
draw attention to the earlier paper by Hernquist \& Weinberg (1992) which showed 
that a rotating, imposed bar in the disk could flatten the cusp also through 
dynamical friction.
\end{itemize}

\noindent
This paper explores and clarifies the possible role played by bars.  I find that 
the mechanism proposed by Weinberg \& Katz (2001) is, in fact, overwhelmed by 
on-going compression leading to a gradual {\it increase\/} in halo density.  The 
possible role of satellites will be explored in a later paper.

Concerns about the simulations and conclusions of Debattista \& Sellwood (1998, 
2000) have been expressed recently.  This paper also addresses the numerical 
issues raised by Weinberg \& Katz (2001).  The criticisms of Valenzuela \& 
Klypin (2002) will be addressed in a future paper.

\section{Mechanisms and Numerical Issues}
Dynamical friction in a perfectly collisionless system (Chandrasekhar 1943) is 
caused by the density wake induced by a massive perturber moving through a 
background of low-mass particles; the gravitational attraction between the wake 
and the perturber is the retarding force.  The standard analysis (Binney \& 
Tremaine 1987) neglects the interactions between the background particles 
themselves, which would tend to increase the concentration of the wake, and 
applies to a steady state in which the introduction of the perturber is ignored. 
 The wake would not form instantly if a perturber were suddenly introduced, but 
would build up over the time scale on which the orbits are deflected.

The analysis becomes more complicated for a bar rotating in a galaxy because 
orbits are periodic.  Repeated encounters with a rotating bar lead to secular 
changes only at resonances (Tremaine \& Weinberg 1984), but friction is still 
expected (Weinberg 1985) because resonant particles generally gain angular 
momentum from the bar.  A bi-symmetric density excess develops in a non- or 
slowly-rotating halo as orbits become trapped or linger ``behind'' the moving 
bar giving rise to the frictional force which slows the bar.

Hernquist \& Weinberg (1992, hereafter HW92) and Weinberg \& Katz (2001, 
hereafter WK01) describe experiments in which the angular momentum transferred 
to the halo through dynamical friction from a driven, imposed bar lowers the 
density of the inner halo substantially.  In the later paper, they suggest that 
simulations of very high quality are needed in order to reproduce the proper 
dynamical friction when the DM has a cusped density profile.  

There is no doubt that simple systems with quasi-harmonic cores are much easier 
to simulate than those with steep density gradients; the large ranges of 
densities and time-scales present in cusped halos obviously demand increased 
numerical care.  But WK01 suggest the difficulties are yet greater because there 
are important resonances between the bar and particles within the cusp where a 
small fraction of the mass resides.  They argue that particle noise can destroy 
resonances if an orbit is scattered frequently and suggest that this delicate 
effect would inhibit the correct evolution in low-quality simulations.

\begin{figure}[t]
\centerline{\psfig{file=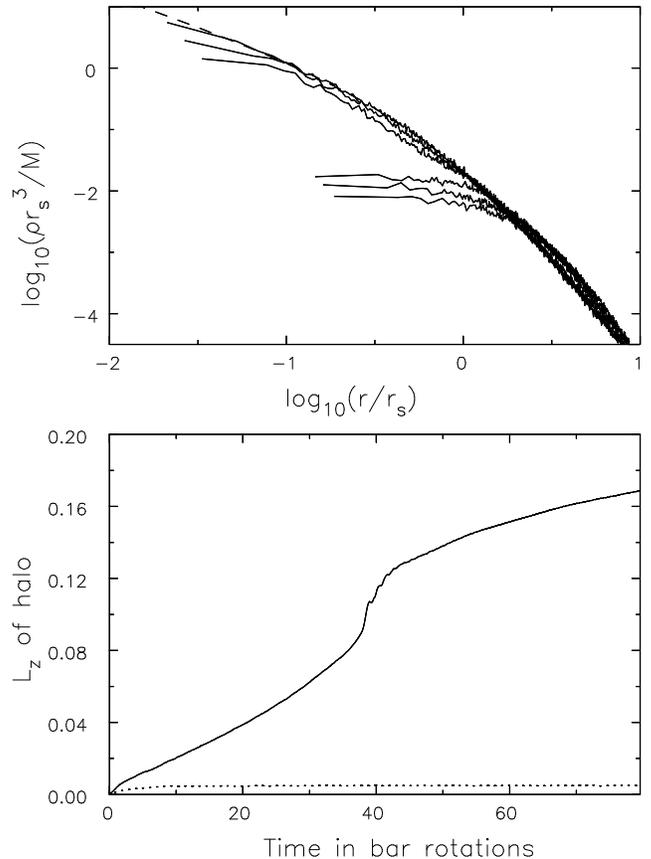,width=\hsize,clip=}}
\caption{\footnotesize Top: The density profile at different times in a model to 
confirm the result from HW92.  The curves are drawn at intervals of 16 bar 
rotations, with the later times having lower central densities.  The dashed 
curve shows the theoretical density profile of the Hernquist model.
Bottom: The angular momentum of the halo as a function of time in the same 
simulation.  The total change of angular momentum is 32 times that contained in 
the bar.  The dotted curve is described in \S3.4.}
\end{figure}

The purpose of the experiments described in \S3 is to determine the numerical 
parameters required to reveal the correct behavior.  I find that cusps can be 
flattened in simulations with relatively modest numbers of particles, but the 
rate of evolution does depend on $N$ in the sense that evolution with too few 
particles can be faster than the correct rate.  This result, while consistent 
with that found by WK01 in their simulations using the SCF method (Hernquist \& 
Ostriker 1992), disagrees with their prediction for collision-dominated methods; 
it would be interesting to test other types of $N$-body code.

\section{Driven bars}
HW92 and WK01 report artificial experiments with bars driven at constant angular 
rates.  In one case, HW92 employed a Hernquist (1990) halo model, which has the 
density profile \be
\rho(r) = {M r_s \over 2\pi r(r_s + r)^3},
\ee having an inner $r^{-1}$ density cusp and an outer slope that is steeper 
than NFW.  Here, $M$ is the total mass of the halo and $r_s$ is the scale radius 
of the profile.  They introduced a rotating quadrupole field of the form \be
\Phi_{\rm bar} = -{GM_{\rm bar} \over a} {\alpha R_*^2 \over (\beta^2 + 
R_*^2)^{5/2}} \sin^2\theta e^{2i(\phi-\Omega_pt)},
\ee where $R_* = R/a$, $R$ is the cylindrical radius, $(r,\theta,\phi)$ are the 
usual spherical polar coordinates, and $M_{\rm bar}$ and $\Omega_p$ are the mass 
and pattern speed of the bar.  This expression, which behaves properly as a 
quadrupole for both small and large $r$, was derived from a fit to the 
analytical field of an inhomogeneous ($n=2$) Ferrers bar with axes $a:b:c = 
1:0.5:0.1$ for which the best fit dimensionless parameters are: $\alpha \simeq 
0.1404$ and $\beta \simeq 0.4372$.  Note that they employed only this $m=2$ 
quadrupole term of the bar field and ignored all other terms, including the 
monopole term which they omitted in order to avoid adding mass in the center of 
the equilibrium halo.

HW92 chose the semi-major axis of the bar, $a$, equal to the scale radius, 
$r_s$, of the Hernquist profile, the bar mass to be 30\% of the halo mass 
interior to this radius (\ie\ 7.5\% of the total halo mass) and drove the bar at 
a constant pattern speed which placed corotation also at $r_s$.  They found that 
the density within the cusp decreased gradually at first, but after about 50 bar 
rotations a rather sudden density decrease completely erased the cusp.

Figure 1 shows the result from an experiment as similar as I was able to 
construct to that reported by HW92.  In this case, I used the PM+SH method 
described in Appendix A; the spherical grid had 200 radial shells, included only 
the $l=0 \, \& \, 2$ terms in the self-gravity of the halo, and employed just 
$100\,000$ particles.  Tests with different time steps, numbers of grid shells 
and larger numbers of particles confirm that the results shown are insensitive 
to these parameters. The evolution of the density profile in the top panel 
closely resembles that shown in their Figure 10, while the change of angular 
momentum of the halo, shown in the lower panel, should be compared to their 
Figure 4.  The agreement with their old results is completely satisfactory.

Even though the set-up is artificial, and the angular momentum given to the halo 
exceeds that contained in the imposed bar by much more than an order of 
magnitude, the experiment illustrates that the halo density profile can, in 
principle, be changed substantially by dynamical friction.

\subsection{Parameter tests}
Figure~2 shows the radius containing 1\% of the halo mass, which obviously 
increases as the central density drops, as function of time from this and other 
experiments using the same halo, bar model and (mostly) numerical code.  
Figure~2(a) shows results from four simulations in which the bar length, $a$, 
and its mass were varied, but the axis ratio, pattern speed and all numerical 
parameters were held fixed.  The bar mass was always 30\% of the halo mass 
interior to $r=a$.  The run with $x=a/r_s=1$ is that shown in Figure 1.  The 
behavior when the bar length is reduced is qualitatively similar, but the 
dramatic change occurs later as the driving quadrupole field is made weaker, and 
did not occur, at least for 160 bar rotations, when $x=0.6$.

\begin{figure}[t]
\centerline{\psfig{file=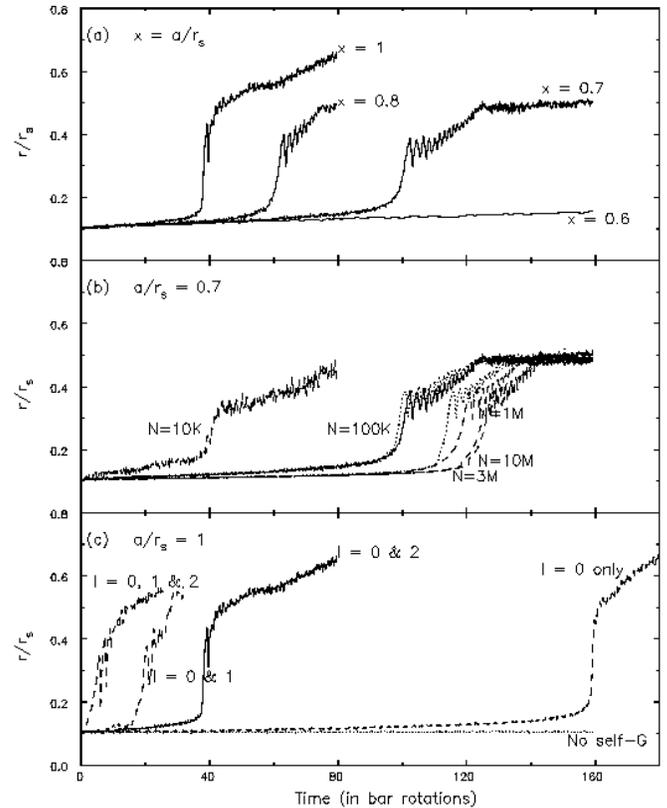,width=\hsize,clip=}}
\caption{\footnotesize Results from many different simulations with a driven, 
imposed bar inside a cuspy Hernquist halo.  (a) The effect of changing the bar 
half-length, $a$, relative to $r_s$, the scale radius of the halo.  (b) The 
effect of changing particle number.  The solid curve for $N=100$K is reproduced 
from (a), the dashed curves are for the same physical parameters but different 
numbers of particles.  The two dotted curves are from runs with the SCF code 
($N=100$K \& $N=1$M) instead of the PM+SH code employed for all others.  (c) The 
effect of changing the order of expansion for the gravitational field.  The 
solid curve is for $x=1$ reproduced from (a) and the dashed lines indicate other 
runs with different numbers of terms in the force calculation.  The dotted line 
shows the effect of ignoring the self-gravity of the halo particles.}
\end{figure}

Figure~2(b) shows the effect of changing the particle number.  The solid curve 
is reproduced from the $x=0.7$ case in (a), and the various dashed curves show 
the results obtained as the particle number is varied while all physical, and 
other numerical, parameters are held fixed.  The trend is clearly that the 
density decrease is delayed as the numerical quality (particle number) is 
improved.  Moreover, the dotted curves show the results for two cases in which 
the field is calculated using the SCF method (Hernquist \& Ostriker 1992) while 
starting from the same initial particle distribution; the differences between 
the results from the PM+SH and SCF codes are relatively minor.

The data for $x=0.7$ in Figure~2(b) hint that the time scale is converging as $N 
\rightarrow 10$M.  This slow convergence for the weaker bar case contrasts with 
the behavior in the strong bar case (not shown); tests with different particle 
numbers when $a=r_s$ ($x=1$) reveal that the rate of evolution is identical when 
$N\gtsim100$K, although it is yet more rapid for $N=10$K -- \ie\ convergence 
when $x=1$ occurs for a particle number some 100 times smaller than for the 
weaker bar case.

Aside from the changing time-scale, the curves in Figure~2(b) are remarkably 
consistent over three orders of magnitude in particle number.  The difference in 
time-scale clearly has little effect on the outcome and the eventual density 
decrease is almost independent of $N$.

Figure 2(c) shows the consequences of changing the number of terms employed to 
determine the field of the halo particles, all for the longest and most massive 
bar in these tests which has $a=r_s$.  The solid line reproduces the $x=1$ 
result from (a) while the dashed curves are for the same particle distribution 
and bar, but employing the indicated expansion terms.  These results show that 
the evolution is accelerated by including $l=1$ terms, and generally that fewer 
terms in the force determination delay the density decrease, which always 
occurs.  Adding $l=3$ and/or $l=4$ terms has little effect; the rate of 
evolution seems to be determined largely by which of the $0 \leq l \leq 2$ terms 
are active.

The horizontal dotted line in Figure~2(c) shows the consequences of dropping 
self-gravity altogether, and integrating a system of test particles in the 
analytic field of the Hernquist halo plus the imposed bar.  While the particle 
distribution gains some angular momentum, the density profile hardly changes 
when self-gravity is suppressed.

\subsection{Cause of the density change}
Resonances are very sharp in a fixed potential well with a steadily rotating 
disturbance.  Once the few particles which start out close to these sharp 
resonances are scattered, or trapped, no more angular momentum can be gained by 
the halo.  The fraction of particles affected by such artificially narrow 
resonances is too small to produce a significant change in the density profile 
in a system of test particles, as shown in Figure~2(c).

The very different behavior of all the models with self-gravity indicates that 
large density changes result from collective effects.  The overall change in the 
mass profile seems to depend only on the bar strength, and not the numerical 
parameters (Figure 2).  However, the rate of evolution does depend on two 
numerical parameters: the number of multipoles included in the calculation of 
halo self-gravity and the number of particles.  I first consider what is 
happening in the large-$N$ limit and then go on to discuss the dependence on 
particle noise.

\subsubsection{Contimuum limit}
The bar scatters halo particles only at resonances in a steady potential.  But 
the orbits of {\it all\/} particles, not just those in resonance, change when 
the gravitational potential well in which they move is altered.  Not only does 
the wake develop, but the overall mass profile of the halo could change, because 
scattered particles gain angular momentum in general.  As the potential well 
changes, more particles are able to become resonant and be scattered, causing 
further density changes.  The process is gradual at first, but the changes 
accelerate as the density disturbance builds.  Even though the process evidently 
runs away, it is not an instability because the continuous supply of angular 
momentum from the bar is needed to drive the on-going re-arrangement of the 
density profile.

The runaway must end once the potential has changed sufficiently that no more 
particles are exposed to the principal resonance.  The subsequent, more gradual 
changes seen in Figure~2 are probably caused by higher order resonances, which 
generally couple less strongly.

\subsubsection{Non-axisymmetric terms}
The rate of evolution depends strongly on which terms contribute to the 
self-gravity of the halo.  It is clear that exchanges are accelerated by 
low-order non-axisymmetric terms, but still occur even when only the monopole 
term is active (Figure 2c).  It should be noted that retrograde particles will 
move closer to the center when given positive $L_z$, while prograde particles 
will move out.  Even though this isotropic model has equal numbers of each 
initially, the net effect is biased and the spherically-averaged mass 
distribution gradually expands as the halo gains angular momentum.

Inclusion of the quadrupole ($l=2$) term has the effect of adding the response 
density to the driving part, which can reasonably be expected to accelerate the 
evolution.  Higher order terms are generally small, and the behavior is little 
affected by them.

The importance of $l=1$ is a surprise, however, since there is no obvious source 
of a dipole field.  I regarded this result with suspicion at first, since White 
(1983) reported that the dipole term caused a numerical artifact under some 
circumstances.  However, tests showed that the rate of evolution with $l=1$ 
active did not depend on such numerical parameters as grid resolution or 
particle number, and persisted even when the initial particle distribution was 
constructed to be perfectly bisymmetric with no net momentum.  The rates of 
evolution both with and without the $l=1$ terms were also reproducible with the 
SCF code.  It is unlikely therefore that it is a numerical artifact.  A dipole 
contribution could possibly arise from an instability related to that discovered 
by Taga \& Iye (1998), although the physical situation is different; a deeper 
investigation is left for future work.

\subsubsection{Particle noise}
The above mechanism may account for the behavior in the large-$N$ limit, but we 
also observe that the runaway occurs earlier in grainier simulations (Figure 
2b).  The number of particles at which the evolution slows to a rate that is 
independent of $N$ is quite modest in some cases, and very large in others, but 
I have found convergence whenever I have tested for it.

The more rapid evolution at low $N$ indicates that fluctuations due to the 
finite number of particles accelerate the density changes.  Some particles which 
would not be resonant in a smooth potential can be scattered into resonance by 
fluctuations (an argument closely analogous to refilling of the loss-cone in AGN 
feeding and cometary dynamics).  Thus the slow-build phase of the density change 
occurs more rapidly, and the runaway phase is reached earlier.  Once the driven 
density changes exceed those arising from noise, the evolution follows the same 
path as in a collisionless model.

Since the runaway is encouraged by fluctuations, one might na\"\i vely expect 
that adding more multipoles at fixed $N$ would hasten the evolution.  It is 
possible that an expansion to very high $l$ would show this effect, but the 
intermediate order ($l=3 \; \& \; 4$) terms appear not to.  (The strong 
dependence on the $l=1 \; \& \; 2$ terms is dynamically real, and probably has 
little to do with particle noise.)

\begin{figure}[t]
\centerline{\psfig{file=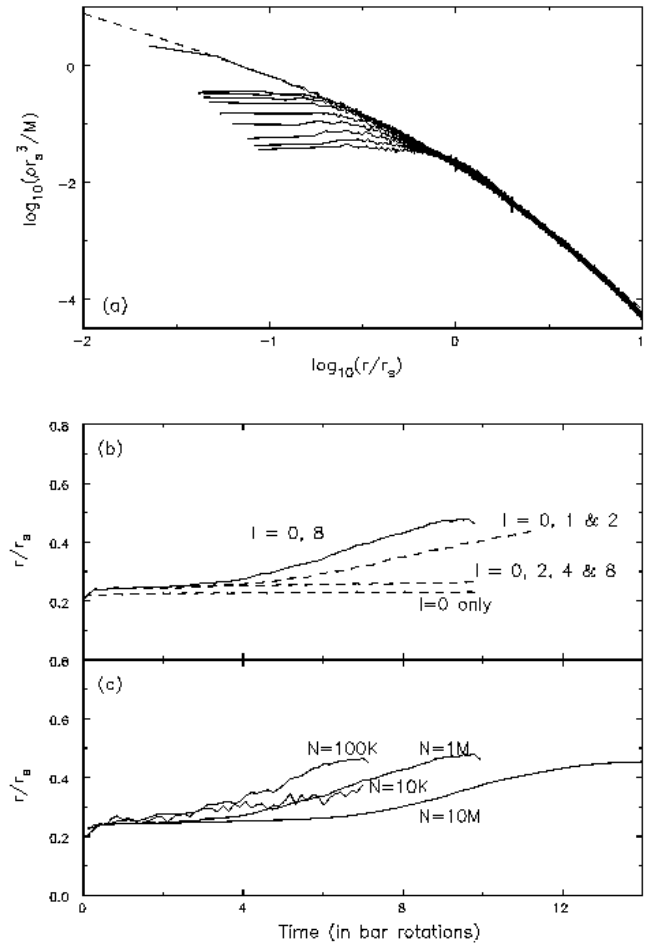,width=\hsize,clip=}}
\caption{\footnotesize (a) The change in density profile of an NFW halo in an 
experiment that approximately reproduces the result reported in WK01.  Curves 
are drawn at intervals of one initial bar rotations in this million particle 
experiment.
(b) The radius containing 1\% of the mass for several experiments with $N=1$M.  
The solid curve shows the result from the simulation shown in (a), which 
includes terms $0 \leq l \leq 8$, and the dashed curves when the force terms 
were more restricted.
(c) The dependence on the number of particles for these experiments with a short 
bar in an NFW halo.  Terms $0 \leq l \leq 8$ are included in each case.}
\end{figure}

\subsection{A shorter bar}
WK01 find a short bar ($a=0.5r_s$) can drive a large density decrease in an NFW 
halo in just 6 bar rotations in a simulation with 4M particles, which seems 
surprising given the trends in my Figure~2.  I was unable to reproduce their 
result at first, but Weinberg (private communication) informed me that they 
employed more terms ($0 \leq l \leq 8$) in the field determination.  When I 
include these additional terms in a simulation, I find the NFW halo, represented 
by 1M particles, does evolve quite rapidly as shown in Figure~3(a).

Figure 3(b) shows that the density change is substantially reduced when odd-$l$ 
terms are omitted in the halo field calculation, and becomes minimal when the 
$l=0$ term only is kept.  These results are further support, of a similar kind, 
for the importance of the $l=1$ terms.  Generally, including terms $l > 2$ has a 
minor effect on the evolution rate, as can be seen in Figure 3(b).

As for the Hernquist halo, I find the rate of evolution generally slows for 
increasing $N$ (Figure 3c), but the run with $N=10$K was so compromised by noise 
that it bucks this trend.  The timescale increases substantially from $N=1$M to 
$N=10$M, indicating that yet larger $N$ is required for convergence when the 
halo is driven by a steadily rotating short bar.

\begin{figure}[t]
\centerline{\psfig{file=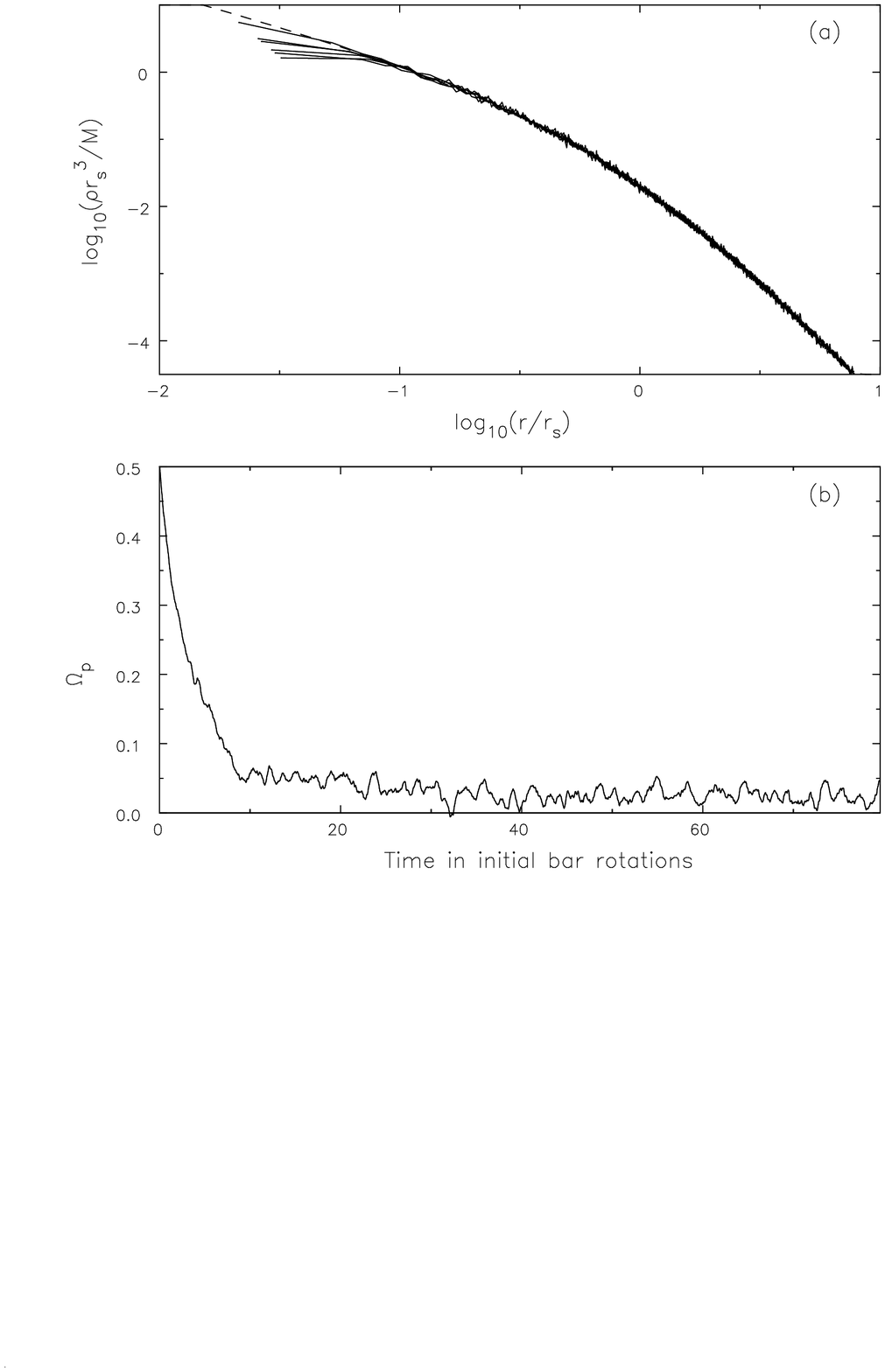,width=\hsize,clip=}}
\caption{\footnotesize (a) The change in density profile of the Hernquist halo 
over the same time interval as shown in Figure~1 when angular momentum is 
conserved.  Curves are drawn at intervals of 16 initial bar rotations.
(b) The pattern speed of the bar which slows as it loses angular momentum.}
\end{figure}

\subsection{Angular momentum conservation}
These simple tests are unphysical, because the bar is imposed and is driven at a 
fixed pattern speed.  As already noted, the total angular momentum given to the 
halo exceeds the angular momentum of the bar by at least an order of magnitude.  
The pattern speed of a rigid bar should therefore decrease rapidly as the bar 
torques up the halo.

Figure 4(b), which is for the same model and parameters as for Figure~1, shows 
that the bar slows rapidly when total angular momentum is conserved, \ie\ as the 
halo gains angular momentum the bar is slowed using the moment of inertia given 
in HW92.  The angular momentum transferred in this case, shown by the dotted 
curve in Figure~1, causes a very minor decrease in the very inner part of the 
halo density (Figure~4a).

But rigid rod-like bars are unrealistic, in part because the effective 
moment-of-inertia of a responsive stellar bar is not simply related to its mass 
distribution.  Fully self-consistent bar halo models, such as described in \S4, 
offer much better tests.

\subsection{Implications}
The trend in Figure~2(b) is that lower noise {\it delays\/} the density 
decrease, as WK01 also found in their SCF simulations.  The better the 
simulation, the harder it becomes to drive a density change suggesting that the 
resonance(s) responsible for the large density change must already be present in 
experiments with quite low $N$.

WK01 argue that 2-body scattering will prevent the correct angular momentum 
exchange with the bar, because an orbit suffers random impulses which destroy 
the important resonances.  However, the empirical evidence from Figures~2 \& 3 
is that only the rate of evolution changes, while the overall angular momentum 
exchange is unaffected by the graininess.  The reason their argument fails for 
these simulations is that scattering is due to large-scale potential variations, 
and is not of the direct 2-body kind.  Any method to determine the gravitational 
field which smooths small-scale density variations suppresses direct 2-body 
scattering, causing orbits to approximate those in a collisionless system, 
albeit in a global potential field which retains some stochastic jitter.  It 
seems reasonable that approximately correct angular momentum exchange between 
particles and a bar could survive at low order resonances under these 
conditions.  It will be interesting to test other $N$-body methods to determine 
whether any behave as WK01 predict.

These wholly artificial experiments, which showed such an encouraging reduction 
in the central density, are misleading for many reasons.  Rigid bars driven at 
constant angular rates imply a huge external source of angular momentum and the 
density reduction is tiny when such a source is absent (Fig.~4).  Also the 
choice of a constant pattern speed is inadvisable, because obtaining the correct 
response when resonances are narrow requires more particles, as WK01 correctly 
point out.

It seems that the severe numerical difficulties presented by this totally 
artifical situation misled WK01 into arguing that simulations of extremely high 
quality are needed to get the physics right.  In a sense, they are correct, 
since their artifical problem is so difficult, but needlessly so.  
Time-dependence of any aspect of the driving field, such as its strength or 
pattern speed, would broaden the resonances and expose a much larger fraction of 
particles to resonant interactions.  Such experiments would be much easier, and 
quite modest simulations should quickly converge to the correct outcome.

\begin{figure}[t]
\centerline{\psfig{file=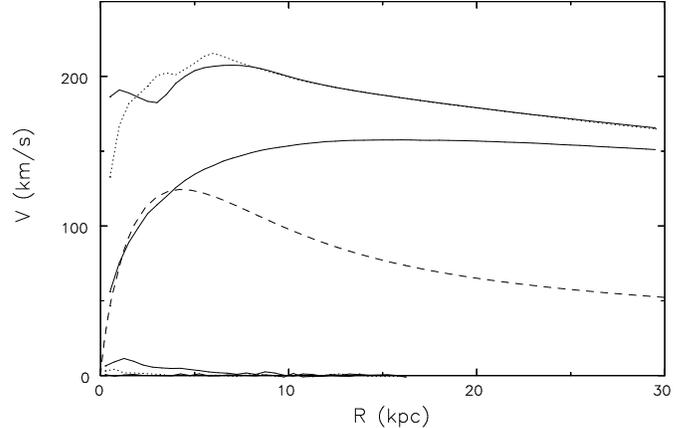,width=\hsize,clip=}}
\caption{\footnotesize Net rotation curves at three different times: the initial 
halo only (solid), at the end of disk growth (dotted), and after the bar forms 
(solid).  Also shown is the contribution of the axisymmetric disk (dashed) and 
very small net rotation acquired by the halo by the end.}
\end{figure}

\begin{figure*}[t]
\centerline{\psfig{file=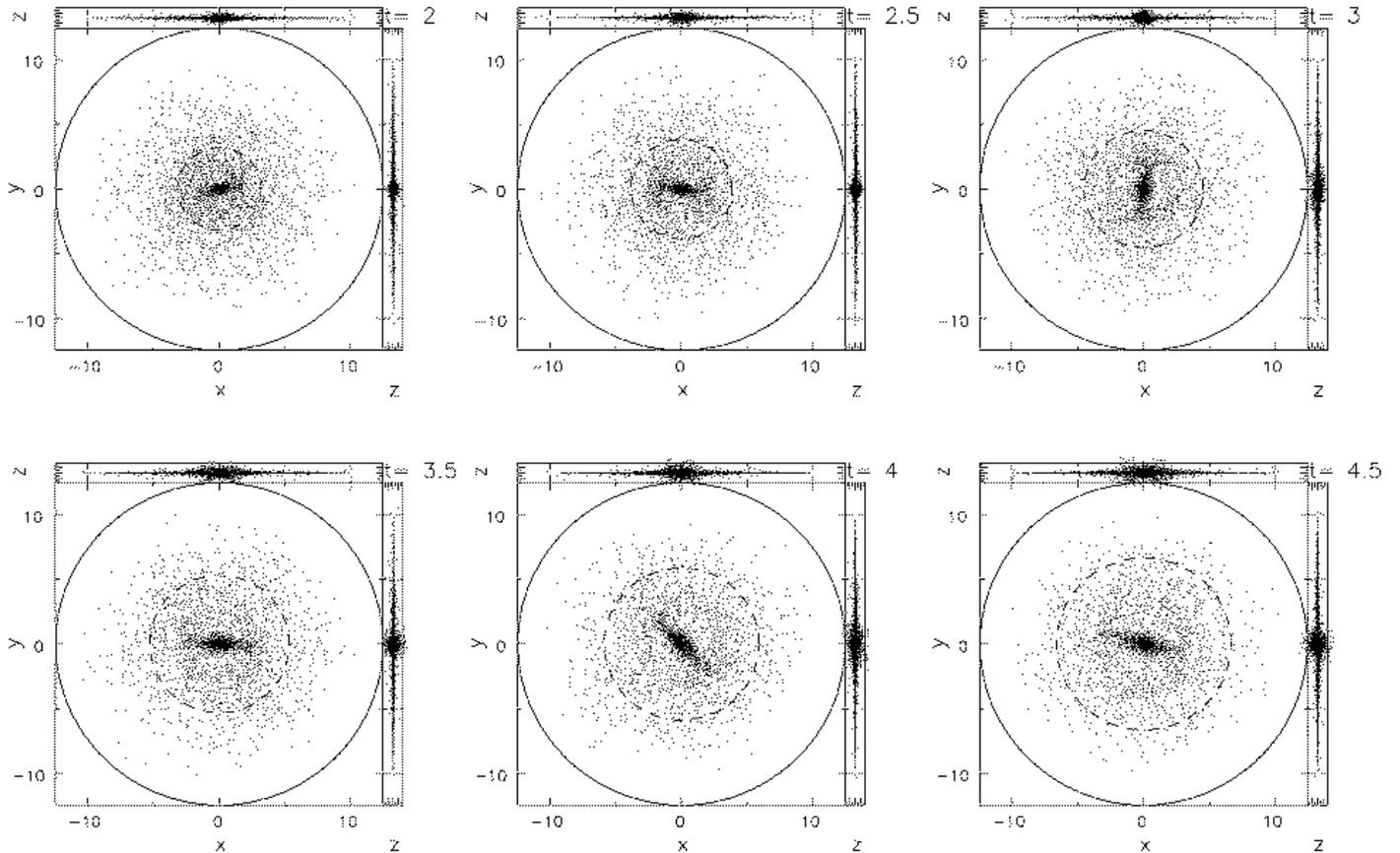,width=\hsize,angle=270,clip=}}
\caption{\footnotesize The later part of the evolution of the disk component.  
Only one particle in 200 is plotted.  The dashed circles show the instantaneous 
radii of corotation.  Times are in Gyr and distances in kpc.}
\end{figure*}

A density runaway in the inner halo may yet be possible, although a very large 
amount of angular momentum must be delivered to the cusp to sustain it.  El-Zant 
\etal\ (2001) suggest that orbiting mass clumps could have this effect on the 
inner density profile, but do not use a self-consistent $N$-body method.  An 
orbiting satellite has much greater specific angular momentum than does a bar, 
and the halo density profile could be flattened significantly if coupling to the 
cusp can be arranged.  Weinberg \& Katz (private communication) have some 
preliminary results, but further experimentation is needed to determine how 
effectively a satellite can couple to the cusp.

\section{Fully self-consistent models}
Here I describe fully self-consistent models designed to investigate whether the 
halo density cusp could be erased through interactions with a responsive bar.  
These experiments required a new code, described in Appendix B, which is 
designed specifically to study isolated, near-equilibrium, disk+halo galaxy 
models.  The method has the dynamic range to resolve collective modes in a 
realistic disk at the center of an extensive halo, while retaining the 
computational efficiency of a grid code.

\subsection{Results}
Creating fully-self consistent disk-halo equilibria is not straightforward and a 
number of increasingly sophisticated methods have been proposed (Hernquist 1993; 
Kuijken \& Dubinski 1995; Debattista \& Sellwood 2000).  Since I wish to 
construct a model with a halo density profile resembling a compressed-NFW form, 
it is easier simply to grow the disk adiabatically within the halo.

I selected an isotropic Hernquist model for the halo (eq.~1), which is similar 
to NFW in the inner parts but the density declines more steeply at large radii.  
I grow an exponential disk within the responsive halo by gradually adding 
particles on locally-determined circular orbits until the disk mass reaches the 
desired value, after which the model evolves freely.  Provided the disk mass 
increases adiabatically, the precise procedure for creating the final model is 
immaterial (\eg\ Binney \& Tremaine 1987, \S3.6), and guarantees an equilibrium 
starting point.  I have run several calculations with different numerical 
parameters and different disk growth-rates to verify that the results are 
insensitive to the specific choices.

Figure 5 shows the rotation curve of one model, computed from the azimuthally 
averaged central attraction, at three different times: the initial halo, after 
disk growth and at the end of the calculation.  The Figure also shows the 
theoretical contribution from a smooth disk (dashed curve) -- in practice the 
disk mass distribution has evolved by the time it reaches full mass; note also 
that the initial halo is compressed by the addition of the disk mass.  While 
clearly sub-maximal, the final disk is significantly self-gravitating; its mass 
is 1/20 of that of the halo and the exponential disk length scale length $R_D = 
r_s/8$.  I have scaled the model, so as to resemble an NFW halo of concentration 
index $\sim10$: $R_D = 2\;$kpc and $V_{\rm flat} = 200\;$km/s.

\begin{table}[t]
\caption{Numerical parameters used in the fully self-consistent simulations}
\label{params}
\begin{tabular}{@{}lrr}
                   & Cylindrical grid       & Spherical grid \\
\hline
Grid size          & $(N_R,N_\phi,N_z)\quad$ \\
                   &  $ = (81,128,125)$     & $N_r = 500$ \\
Angular compnts    & $0\leq m \leq 8$       & $0 \leq l \leq 4$ \\
Outer radius       & 12.4 kpc               & 320 kpc \\
$z$-spacing        & 12.5 pc \\
Softening length   & 25 pc  \\
Largest $N$        & 1M                     & 20M \\
Shortest time step & 0.5 Myr                & 0.5 Myr \\
\hline
\end{tabular}
\end{table}

I have run several models with very similar physical parameters, but with the 
number of halo particles ranging from $2 \times 10^5 \leq N \leq 2 \times 10^7$; 
other numerical parameters are summarized in Table~1.  Since I expanded the 
density on the spherical grid up to the $l_{\rm max}=4$ terms only, the halo 
particles feel only a rather fuzzy disk field when near the outer disk, but the 
full field of the bar is well represented.  Tests with changing $l_{\rm max}$ 
showed that the bar-halo torque is almost entirely due to the large-scale $l=2$ 
terms and the overall behavior is not all sensitive to increased expansion 
order.

Results reported here are from the model with 20M halo particles and 1M disk 
particles; I do not find the evolution is significantly different for either one 
fourth or one twentieth this number.  The model with just 200K halo particles 
and 10K disk particles did differ, however; it did not form a bar in the disk at 
all, presumably because noise heated the disk too much.

\begin{figure}[t]
\centerline{\psfig{file=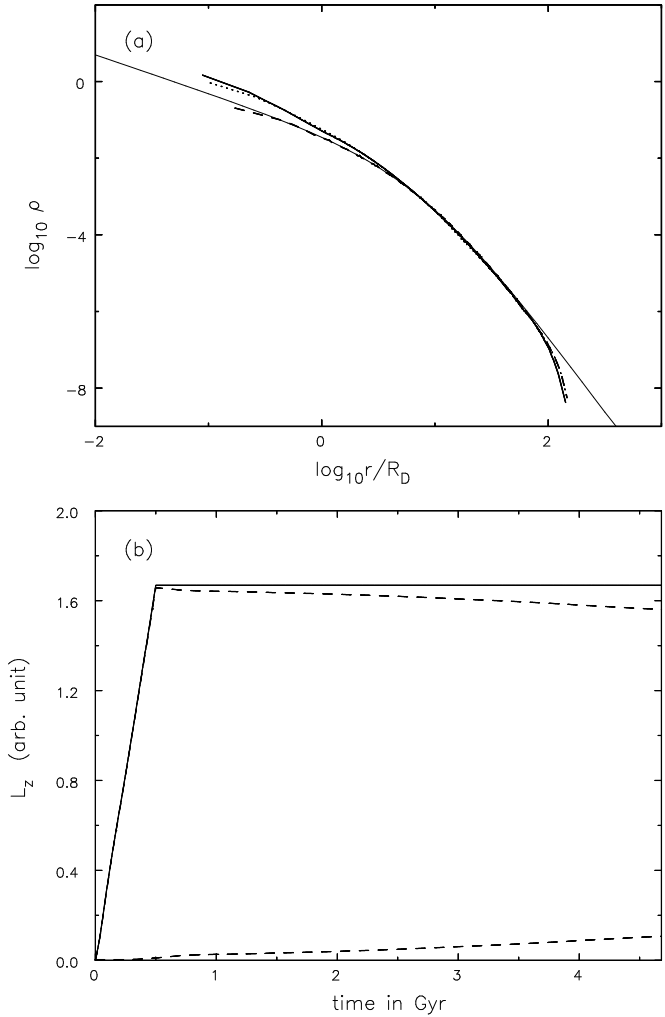,width=\hsize,clip=}}
\caption{\footnotesize (a) The density profile of the halo in the simulation 
described in \S4.1 at three different times.  The smooth curve shows the 
theoretical Hernquist profile, while the dashed curve is measured from the 
particles at the start.  The dotted curve shows the density profile after disk 
growth (0.5~Gyr) and the solid curve shows the profile at the end of the 
simulation ($\sim 5\;$Gyr).  The scale radius of the initial halo is $r_s=8R_D$, 
and its truncation at $r=160R_D$.
(b) The total angular momentum of the model (solid curve) and the separate 
angular momenta of the disk and halo (dashed curves).  The disk grows over the 
first 0.5~Gyr and then slowly loses angular momentum to the halo, which has none 
at first.}
\end{figure}

The mass of the exponential disk grows linearly over a period of 0.5~Gyr.  As it 
nears its final value, the disk develops strong spirals, which are multi-armed 
at first, but are later dominated by $m=3$ and some $m=2$.  A bar forms after 
about 1.2~Gyr; linear theory (Toomre 1981) would predict that a disk in a halo 
of this type is stable to bar-modes, a prediction that has been verified in 
extremely high-quality simulations (Sellwood \& Evans 2001).  However, 
swing-amplified particle noise, which produces the spirals, leads to a bar 
through (non-linear) orbit trapping (\eg\ Sellwood 1989) even for quite large 
particle numbers when, as here, the disk is highly responsive.

Figure 6 shows the later part of the evolution of the disk component.  The 
half-length of the bar, $a_{\rm B}$, was about 2~kpc, or only about 1/8 of the 
halo length scale, at first, but by 4.5~Gyr it had increased to about 3.5~kpc.  
The bar also slows down, as can be deduced from the corotation circle drawn by 
the dashed lines, to the extent that $a_{\rm B}$ is only about half the distance 
to corotation in the later stages of evolution.  Such a slow bar is inconsistent 
with those observed (Aguerri \etal\ 2002).

\begin{figure}[t]
\centerline{\psfig{file=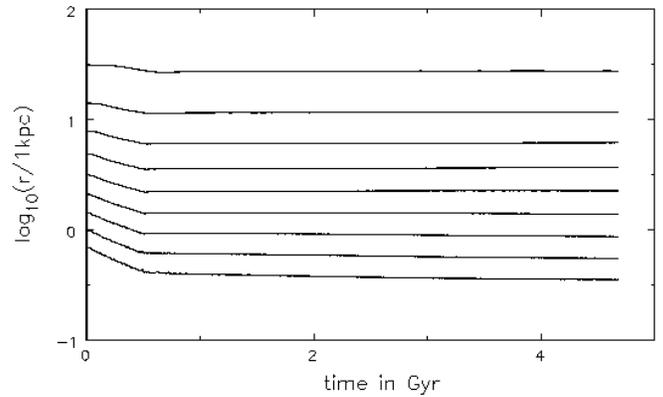,width=\hsize,clip=}}
\caption{\footnotesize Radii enclosing various mass fractions as a function of 
time.  Each line represents an increase of a factor 2 in enclosed mass, from 
0.2\% to 51.2\% of the total Hernquist halo, which was truncated at 320~kpc.}
\end{figure}

Figure 7 shows the density profile of the halo at three times, together with the 
analytic density profile of the initial halo.  The process of disk growth causes 
the halo to contract, of course, and the inner density profile steepens in the 
first 0.5~Gyr.  The formation of the bar within the disk, which concentrates the 
disk mass, compresses the inner halo slightly more, giving the solid curve which 
stays constant until the end of the calculation at $t\simeq 5\;$Gyr.  Dynamical 
friction transfers angular momentum to the halo (Figure 7b), but the total 
amount is too small to reduce the halo density significantly.  In fact, Figure~8 
shows that the entire evolution is always towards a more compressed halo, caused 
by disk growth followed by on-going mass rearrangement within the disk.  
Whatever density reduction of the inner halo that is being caused by bar 
scattering is being overwhelmed by the extra compression as the disk mass 
becomes more concentrated.


\subsection{Discussion}
Results from these models suggest that bars in fully-self-consistent disks with 
NFW-like halo models are too short ($0.125 \ltsim a_{\rm B}/r_s \ltsim 0.25$) to 
erase the cusp by the method proposed by WK01.  The bar length is determined by 
disk dynamics and is not, therefore, an adjustable parameter.  More extensive 
halos with larger $r_s$ or lower $c$ might possibly allow longer bars to form, 
but I have so far failed to construct a realistic galaxy model (\ie\ one having 
a quasi-flat rotation curve) which allows a large-enough bar to form in the disk 
that could conceivably erase the halo cusp.


\section{Conclusions}
The results presented here indicate that the inconsistency between the cusped 
halos predicted from CDM models and the observed cores in galaxies cannot be 
removed by bar interactions.  The suggestion by Weinberg \& Katz (2001) that 
this might be possible was based on preliminary tests with large imposed bars 
with an unrealistic store of angular momentum.

Self-consistent bars do not contain enough angular momentum to change the halo 
density profile.  Relatively short bars form in cuspy halos through disk 
``instabilities'' on a scale controlled by the steeply rising rotation curve.  
The small bar exerts a correspondingly weak torque, and its small moment of 
inertia implies that little angular momentum is available to be transferred to 
the halo.\footnote{Larger bars might form in low-resolution simulations, because 
the steep rise in the inner rotation curve is inadequately resolved, but such 
simulations cannot address the issues raised here.}

Additionally, as the bar forms, grows, and loses angular momentum, the mass 
distribution within the disk becomes more concentrated.  The response of the 
halo is further compression which overwhelms any tendency for the inner halo 
density to decrease through its mild angular momentum gain.

I have also examined a number of numerical issues raised by WK01.  The 
experiments they chose to illustrate the difficulties they describe were 
needlessly delicate because they contain extremely sharp resonances.  More 
realistic tests in which the pattern speed declines as the bar loses angular 
momentum, have broader resonances and are therefore much easier.  Nevertheless, 
I find that the density cusp can be flattened {\it more readily\/} in 
experiments with fewer particles in agreement with their findings with the SCF 
code.  Even though the rate of evolution can be $N$-dependent, these simulations 
always converged to physically reasonable behavior that was independent of $N$.  
Conditions which require extremely high $N$ are squarely in the needlessly 
difficult category.  This does not, of course, imply that all $N$-body 
simulations can be trusted; those which are sloppily set up or fail to 
approximate the collisionless limit should always be regarded with suspicion.

\acknowledgments
Martin Weinberg, who continues to disagree with some of my interpretations, has 
helped considerably with my understanding of the issues in this paper and has 
patiently borne an extensive e-mail correspondence and numerous discussions.  
Thanks are due to Victor Debattista for many helpful comments and for suggesting 
improvements to an early draft of the paper.  Comments from Arthur Kosowsky and 
Eric Barnes were also helpful.  This work was supported by NASA SARA grant NAG 
5-10110 and by NSF grant AST-0098282.

\appendix
\section{A. Spherical grid}

\def\pmb#1{\setbox0=\hbox{$#1$}%
  \kern-0.25em\copy0\kern-\wd0
  \kern.05em\copy0\kern-\wd0
  \kern-0.025em\raise.0433em\box0}
\def\br{\;\pmb{\mit r}}

Several authors (van Albada \& van Gorkom 1977; van Albada 1982; Villumsen 1982; 
White 1983; McGlynn 1984) have previously described $N$-body codes based on an 
expansion in surface harmonics.  The potential at a field point 
$(r,\theta,\phi)$ due to a number of point masses at 
$(r_j^\prime,\theta_j^\prime,\phi_j^\prime)$ is the real part of 
\begin{eqnarray}
 \Phi(\br) = & -G\displaystyle \sum_{l=0}^\infty \sum_{m=0}^l (2-\delta_{m0}) 
{(l-m)! \over (l+m)!}
\times \nonumber \\
& \qquad P_l^m(\cos\theta) e^{im\phi} \displaystyle \left[ {A_{lm} \over 
r^{l+1}} + r^l
B_{lm} \right].
\end{eqnarray} where \begin{eqnarray}
A_{lm} \equiv & \displaystyle \sum_{j, \rm int} M_j P_l^m
(\cos\theta_j^\prime)e^{-im\phi_j^\prime} r_j^{\prime l} \nonumber \\
B_{lm} \equiv & \displaystyle \sum_{j, \rm ext} M_j P_l^m
(\cos\theta_j^\prime)e^{-im\phi_j^\prime} r_j^{\prime -(l+1)}.
\end{eqnarray} The summation for $A_{lm}$ is over all particles for which 
$r_j^\prime < r$, while all those with $r_j^\prime > r$ are included in that for 
$B_{lm}$.

The two principal problems to be faced when using this formula for the forces in 
an $N$-body code are: (1) The errors in the calculated forces between a pair of 
particles at similar radii can be very large when the summations are truncated 
at a finite $l_{\rm max}$.  Increasing $l_{\rm max}$ reduces the radial extent 
of the region where the error is large, but also causes the errors over this 
shrinking region to worsen (just as Gibbs's phenomenon in Fourier series).  (2) 
The forces acting between the pair of particles change discontinuously as they 
cross in radius -- the problem known as ``shell crossings.''  Discontinuous 
changes in acceleration, when combined with a finite time-step size, cause 
particles to gain or lose energy for numerical reasons only.  Villumsen (1982) 
and White (1983) introduce a softening parameter into the expressions for the 
potential, which smooths the forces but also creates a subtle artifact (White 
1983).  An alternative strategy (van Albada 1982), is to introduce a grid in all 
three coordinates.  I have adopted the intermediate approach advocated by 
McGlynn (1984).

I introduce a 1-D radial mesh on which the coefficients $A_{lm}$ and $B_{lm}$ 
are tabulated, but retain the exact angular dependence, and therefore denote 
this method as PM+SH (standing for particle-mesh + surface harmonics).  
Concentric spheres at the set of radii $\{r_k\}$, $k=0,n$ with $r_0=0$, divide 
the computation volume into onion-like shells.  I improve slightly on McGlynn's 
method by interpolating between grid radii in the spirit of the cloud-in-cell 
procedure widely used in PM methods as follows:
For a particle at radius $r_j^\prime $, I locate the nearest $r_k$ (which can be 
either interior or exterior to $r_j^\prime $).  I then give that particle the 
finite radial extent $\delta r_j^\prime  = {1\over2}(r_{k+1} - r_{k-1})$, 
centred on $r_j^\prime $.  Thus fractions \be
w_{j,1} = {1\over2} - {r_j^\prime -r_k \over \delta r_j^\prime } 
\qquad\hbox{and}\qquad
w_{j,2} = {1\over2} + {r_j^\prime -r_k \over \delta r_j^\prime }
\ee of the mass of particle $j$ lie interior and exterior respectively to $r_k$. 
 Exceptions to this rule occur at $r_0$, where the entire mass of the particle 
is deemed to be exterior to the centre, and at $r_n$, where the entire mass of 
the particle is deemed to be interior to the outer edge.

The first step is to form the partial sum of the contributions from each 
particle fragment to the interior and exterior terms on each of its two 
neighbouring grid points, for each $(l,m)$: \begin{eqnarray}
\alpha_{lm}(k) & = \sum M_j w_{j,\nu } P_l^m (\cos\theta_j^\prime) 
e^{-im\phi_j^\prime} \displaystyle {1 \over r_k} \left( r_j^\prime \over r_k 
\right)^l \cr
\beta_{lm}(k) & = \sum M_j w_{j,\nu } P_l^m (\cos\theta_j^\prime) 
e^{-im\phi_j^\prime} \displaystyle {1 \over r_j^\prime} \left( {r_k \over 
r_j^\prime} \right)^l, \cr
\end{eqnarray} where the summation for $\alpha_{lm}(k)$ [$\beta_{lm}(k)$] is 
over particle fragments, $\nu =1,2$, in the shell immediately interior 
[exterior] to $r_k$, but no further than $r_{k-1}$ [$r_{k+1}$] \ie\ every 
particle contributes to just four complex coefficients.  (There are no 
contributions to $\alpha_{lm}(0)$ since no mass can be interior to $r=0$.)  Note 
that $\alpha_{lm}(k)$ and $\beta_{lm}(k)$ include the factors $r_k^{-(l+1)}$ and 
$r_k^l$ in the particle's contribution to the coefficients $A_{lm}(k)$ and 
$B_{lm}(k)$ on sphere $k$.  Since the ratio $r_k/r^\prime_j$ is close to unity, 
including these factors avoids multiplication (division) by large (small) 
numbers raised to high powers, as recommended by van Albada \& van Gorkom 
(1977).  (The exception for $r_0$ causes no difficulties since only the monopole 
term contributes at the center -- the angular terms have no meaning.)

The next step is to sum over grid points so that a single coefficient on each 
grid point includes contributions from all the interior and exterior mass.  This 
is a recursive sweep both inwards and outwards combining the previously obtained 
coefficients as follows: \begin{eqnarray}
A_{lm}(k) & = \alpha_{lm}(k) + A_{lm}(k-1) \displaystyle \left( {r_{k-1} \over 
r_k} \right)^{l+1} \nonumber \\
B_{lm}(k) & = \beta_{lm}(k) + B_{lm}(k+1) \displaystyle \left( {r_k \over 
r_{k+1}} \right)^l.
\end{eqnarray} Note, this is the only step in the method for which the 
computation time depends on the number of grid radial grid points employed.  
Since there should be many fewer shells than particles, this single sweep in 
each direction represents a tiny overhead.

The potential is now defined on each sphere $r_k$ and the acceleration vector 
can be computed as the negative gradient wrt the field coordinates.  The 
expressions for the derivatives, though tedious, are straightforward.  For the 
potential and acceleration components at general field points, where the 
particles are located, I interpolate linearly between shells for the radially 
dependent parts, but evaluate the angular part at the precise angular direction 
of the particle.

The source mass distribution is smoothed because the point mass particles are 
blurred both radially, by the grid and mass assignement scheme, and azimuthally, 
by the truncations at $l_{\rm max}$.  The amount of radial smoothing is 
determined by the radial grid spacing.  Forces acting on each particle change 
continuously as the particle crosses the grid shells, although the gradient can 
be steep for large $l_{\rm max}$.  Particles experience forces from others in 
the same shell (even themselves), but the relative contribution is small when 
$N$ is large, because each particle has little mass, and the shells are thin, 
suggesting that the number of radial mesh points should vary with number of 
particles.

The radii of the grid points are completely arbitrary, but it is desirable to 
space them unevenly in order to obtain higher resolution near the center where 
the density is highest.  I tried the strategy of spacing the shells so as to 
enclose precisely equal mass fractions, but find that the following rule leads 
to better results:
\be
r_k = e^{\gamma k} - 1, \qquad\hbox{with}\qquad \gamma = {\ln(r_n+1) \over n}
\ee
where $r_n$ is the desired radius of the outermost shell.

The overall computation time of the method scales almost perfectly as $N$ times 
the number of azimuthal terms $(l_{\rm max}+1)(l_{\rm max}+2)/2$, because step 
(A5) is negligible.  The code can be very efficiently adapted for parallel 
architectures; the best strategy is for each processor to work with a fraction 
of the particles.  Once the partial sums (A4) have been formed on each, the only 
numbers which need to be communicated between processors are the rather small 
number of coefficients $\alpha_{lm}(k)$ and $\beta_{lm}(k)$.

\section{B. Hybrid grid $N$-body method}
Particle-mesh (grid) methods are the basis of the most efficient $N$-body codes 
(Sellwood 1997), and are therefore to be preferred for applications where they 
are viable.  The traditional workhorse, the Cartesian grid, is barely adequate 
to resolve disk dynamics (Sellwood \& Merritt 1994) and requires the halo to 
have a very limited extent and large quasi-uniform core (Debattista \& Sellwood 
2000).  Adaptive grids (\eg\ Klypin \etal\ 1999) are able to improve resolution 
wherever the particle density is high, but these locations are known and 
unchanging in our problem, and a fixed grid which concentrates resolution where 
it is needed will be more efficient.

No single fixed grid so far described has the ability resolve a thin disk, 
embedded within an extensive halo having a steep density gradient, however.  
Cylindrical polar grids (Pfenniger \& Freidli 1991; Sellwood \& Valluri 1997) 
offer better disk resolution than do Cartesian grids, but the number of grid 
planes becomes unmanageably large if the same grid must encompass an extensive 
halo, while uneven plane spacing is hopelessly inefficient.  Spherical grids 
(Appendix A) on the other hand, are ideally suited to a near spherical mass 
distributions with a steep density gradient, but are unable to resolve the 
vertical structure of a disk component unless expanded to very high $l_{\rm 
max}$ which is both expensive and undesirable.

I have therefore developed a hybrid PM scheme in which the self-gravity of the 
disk is computed on a high-resolution cylindrical polar grid while that of the 
halo is computed using the PM+SH method.  As there is no restriction on the 
relative extents of the two grids, the initial disk can almost fill the 
cylindrical polar grid while the spherical grid has a radius many (here $\sim 
25$) times larger.

Clearly, all particles must feel forces from all others at every step.  Forces 
acting on disk particles within the smaller grid can be computed by 
interpolation on both grids, but the field of the particles on the smaller grid 
cannot be extended outside that grid.  I therefore define a second spherical 
grid, identical to the first, on which I calculate the forces from the disk to 
be added to the self-consistent halo field.  Forces from the inner disk, bar and 
bulge are well represented by the spherical grid, even though the density 
distribution on the spherical grid is generally expanded to rather low $l_{\rm 
max}$.  The strong vertical forces near the highly flattened outer disk are not 
well represented by the spherical grid, but that is the main purpose of the 
high-resolution small grid.

In order avoid repeated discontinuous changes in forces for particles whose 
orbits take them into and out of the smaller grid volume, I label all halo 
particles as belonging to the larger grid, and use only accelerations determined 
for that grid wherever they may be.  Thus, the halo particles feel only a rather 
fuzzy disk field when near the outer part of the disk; tests with increasing 
$l_{\rm max}$ confirm that the overall dynamics is not significantly affected by 
this approximation.  Disk particles which drift out of the smaller grid, are 
also relabelled as belonging to the larger grid for the rest of the calculation. 
 If strong self-gravity of the outer disk is ever thought to be necessary as the 
disk spreads beyond the inner grid edge, it would be relatively inexpensive to 
increase the size of the inner grid.

The code allows particles to have individual masses, for their motion to be 
integrated using a range of time steps for greater efficiency, and has been 
implemented using the message-passing-interface (MPI) to enable its use on 
parallel machines with high efficiency.

\end{document}